\documentstyle[12pt,epsf]{article}

\newcommand{\ra}{\rightarrow}
\newcommand{\be}{\begin{equation}}
\newcommand{\ee}{\end{equation}}
\newcommand{\bea}{\begin{eqnarray}}
\newcommand{\eea}{\end{eqnarray}}
\newcommand{\ci}{\cite}
\newcommand{\bi}{\bibitem}
\newcommand{\nono}{\nonumber \\}

\newcommand{\s}{\sigma}

\newcommand{\bfnabla}{\mbox{\boldmath$\nabla$}}

\newcommand{\rr}{\vec{\bf{r}}}

\def\dal{\,\lower0.3ex\vbox{\hrule\hbox{\vrule\kern2pt\vbox{\kern4pt\kern4pt}
\kern2pt\vrule}\hrule}\,}

\def\s{\sigma}

\begin{document}

\title{\sl Single and double slit scattering of wavepackets}
\vspace{1 true cm}
\author{G. K\"albermann$^*$
\\Soil and Water dept., Faculty of
Agriculture, Rehovot 76100, Israel}
\maketitle

\begin{abstract}

The scattering of wavepackets from a single slit and a double slit
with the Schr\"odinger equation, 
is studied numerically and analytically.

The phenomenon of diffraction of wavepackets in space and time in 
the backward region, previously found for barriers and wells, is
encountered here also.

A new phenomenon of forward diffraction that occurs only 
for packets thinner than the slit, or slits, is calculated
numerically as well as, in a theoretical approximation to the
problem. 
This diffraction occurs at the opposite end of the
usual diffraction phenomenon with monochromatic waves.
\end{abstract}


$^*${\sl e-mail address: hope@vms.huji.ac.il}

\newpage

\section{\sl Introduction}

In \ci{k1,k2,k3,k4}, the 
phenomenon of wavepacket diffraction in space and time was
described, both numerically in one and two dimensions \ci{k1,k2},
 as well as analytically in one and three dimensions \ci{k3}. 
The phenomenon occurs in wavepacket potential scattering for the
 nonrelativistic, Schr\"odinger equation and for the relativistic Dirac
equation.

The main feature of the effect, consists in the production of
a multiple peak structure that travels in space. This pattern was
interpreted in terms of the 
interference between the incoming, spreading wavepacket, and, the scattered 
 wave.
The patterns are produced by a time independent potential in the
backward direction, in one dimension, and, at large angles, in three
dimensions.

The multiple-peak wave train exists for all packets, but, it does not
decay only for
packets that are initially thinner than $\sqrt{\frac{w}{q}}$, where
 {\sl w} is a typical potential range or well width and {\sl q}
 is the incoming average packet momentum.
For packets that do not obey this condition, the peak structure
eventually merges into a single peak.

The effect was named:  {\sl Wavepacket diffraction in space and time}. 
(See \ci{k3} for details)

The present work addresses the paradigm of wave phenomena: 
Diffraction from a single slit and a double slit, this time with
wavepackets.
The outcome of this study is twofold.
 In the backward region, the phenomenon of wavepacket diffraction will
emerge, while, in the forward region a new diffraction phenomenon
will be found.

Diffraction phenomena from slits with electromagnetic waves \ci{Born}, as
well as matter waves diffraction from slits \ci{cow},\ci{zei1}, are 
treated by means of plane monochromatic waves.
The Kirchhoff approximation \ci{jackson}, 
together with Green's theorem and Dirichlet boundary
conditions, suffice to reproduce the alternating
intensity structure on a screen or detector, 
the well known Fresnel and Fraunhofer patterns in the forward region.
In either case, the treatment of wavepackets is absent from
the literature.

A smooth 
packet may be seen as made of a continuous spectrum of frequencies.
From Huygens principle, it is expected that, destructive interference between
the various monochromatic components will generate a peak only in the 
forward direction. The thinner the incoming packet, 
the broader the energy-momentum spectrum. Therefore, 
a very thin packet should by no means display a diffraction pattern. 
It is perhaps for this, and technological reasons, that there
have not been any 
experimental searches and theoretical efforts to
deal with the diffraction of wavepackets.
There exist studies of electron diffraction with an electron microscope
and of cold neutrons with crystals as reviewed in \ci{zei1}. The
 experiments do not consider the packet structure.

The present work shows numerically and analytically that, 
contrary to the expectations, 
there appear diffraction patterns in the forward direction
for packets, that are thinner initially as 
compared to the slits dimensions.

Section 2 surveys briefly the phenomenon of diffraction of
wavepackets in space and time that prompted this investigation. 
Numerical results for a single slit and a double slit are presented
in this section. The results show that, 
the wave in the backward region for narrow packets, appears as a 
propagating multiple peak wave train, even in the region facing the slit
holes. The theoretical analysis of this phenomenon 
follows directly from the formulae of reference \ci{k3}.
Section 3 shows the results for the forward zone.
The clear distinction between a thin packet and a wide, but finite, packet is
evidenced by the existence or absence of a diffraction pattern.

A full analytical solution to the slit problem, even for plane waves, does not
exist. There exists a treatment of  
wavepacket diffraction from slits with a separable ansatz \ci{zecca}. 
However this assumption is not valid, except 
perhaps at very large distances from the slit, or slits.

As an approximation to the exact analytical solution, 
the method of Green's functions is applied in section 3.
The predicted patterns using the Kirchhoff approximation, 
agree with the numerical results, except for thin packets.

Section 4 describes an improvement over the Kirchhoff approximation.
The approach considers the added effect of the standing waves, 
or cavity modes inside the slit. 
The ensuing diffraction pattern generated by the transient and
cavity modes is called, {\sl cavity mode diffraction.}
For extremely wide packets, resembling plane waves, 
the cavity modes contribution vanishes.

The long time behavior of the the forward scattering is 
depicted using the improved wave function.
This aspect cannot be evaluated numerically at the present time, 
because it requires enormous computer time and memory.
The theoretical predictions show that the pattern of a thin packet 
changes with space and time. 
It becomes extremely complicated as time evolves, but, it 
does not merge into a single peak as would have been expected. 
For extremely wide packets, the predicted pattern is identical to
 the one of a plane wave.

Section 5 summarizes the paper.

\section{\sl Slit diffraction in the backward region}

Plane wave monochromatic waves in quantum mechanics show
diffraction phenomena in time \ci{mosh}, induced by the sudden
opening of a slit, or in space by fixed slits or gratings.
The combined effect of time dependent opening of slits
for plane monochromatic waves produces diffraction patterns in space and
time \ci{zei2}. 
These patterns tend to disappear as time progresses.
Atomic wave diffraction experiments \ci{prl}, have confirmed the predictions
of Moshinsky \ci{mosh}, to be correct.

The phenomenon of diffraction of wavepackets 
in space and time was presented in \ci{k1,k2,k3,k4}.
It consists of a multiple peak traveling structure,
generated by the scattering of initially thin packets from 
a time independent potential, a well or a barrier.

In \ci{k3}, analytical formulae for $t\ra\infty$ were developed
for the diffraction pattern of a scattering event of a wavepacket at
 a well or barrier. 

In one dimension the amplitude of the wave in the backward region reads

\bea\label{asymp}
|\psi(x,t)|&=&2~\sqrt{\frac{2~m~\pi}{t}}~e^{-z}~|sin\bigg(\frac{m~x}{t}(x_0
+2~i~\s^2~q_0)\bigg)|\nono
&=&2~\sqrt{\frac{2~m~\pi}{t}}~e^{-z}~\sqrt{sin^2(\frac{m~x~x_0}{t})+
sinh^2(\frac{2~\s^2~q_0~m~x}{t})}\nono
z&=&\s^2~\bigg(\frac{m^2~(x^2+x_0^2)}{t^2}+q_0^2\bigg)
\eea

where {\sl m} is the mass $x_0$, the location of the
center of the packet at $t~=~0$, $\s$, approximately equal to 
the full width at half maximum of the
packet for a gaussian packet, and, $q_0~=~m~v$, the average 
momentum of the packet.

This expression represents a diffraction pattern that travels in time.
Eq.(\ref{asymp}) was compared to the numerical solution in \ci{k3}, with 
excellent agreement, and without resorting to 
any scale factor adjustment.(Figure 1 in \ci{k3}).

The condition for the pattern to persist, 
derived basically from eq.(\ref{asymp}), was found to be

\bea\label{constraint}
\s<<\sqrt{\frac{w}{q_0}}
\eea

The key element entering the derivation of the formula of eq.(\ref{asymp}), 
was the replacement of the real part of the reflection coefficient 
by its value at threshold, namely {\sl{Re(R)=-1}}. This
replacement is valid for very long times, at which the wild oscillations
in the integrals that determine the reflected wave, 
favor the contributions of very low momentum. 
The value at threshold is independent of the type of well
 or barrier. Hence, the result holds in general.
Depending on the initial width of the packet, 
a receding multiple peak coherent wave train or, a single hump, will appear.
This is the essence of the phenomenon of {\sl wavepacket diffraction in space
and time}.
In three dimensions the traveling diffractive structure may occur at
several angles \ci{k3}.

The present work grew out of the interest in the case of slit diffraction
as a laboratory for the phenomenon of wavepacket diffraction.

This section, deals with the backward region for single and double 
slit scattering, the appearance of
diffraction patterns in the forward region, 
is discussed in section 4.
\vspace{1 cm}

Figure 1 depicts the single slit setup. Physical slits
are usually infinitely long as compared to their width. Therefore
a two-dimensional treatment is relevant.
The impinging wave advances from the left towards the slit. 
The screen that defines the slit, is taken as an impervious
 surface. In actual numerical calculations 
a large value for this repulsive potential was used, typically several orders
of magnitude larger than the average kinetic energy of the packet.
Differing from the the usual treatments of slit diffraction, 
the slit length {\sl 2 a} in fig. 1 will play here a major role.

\begin{figure}
\epsffile{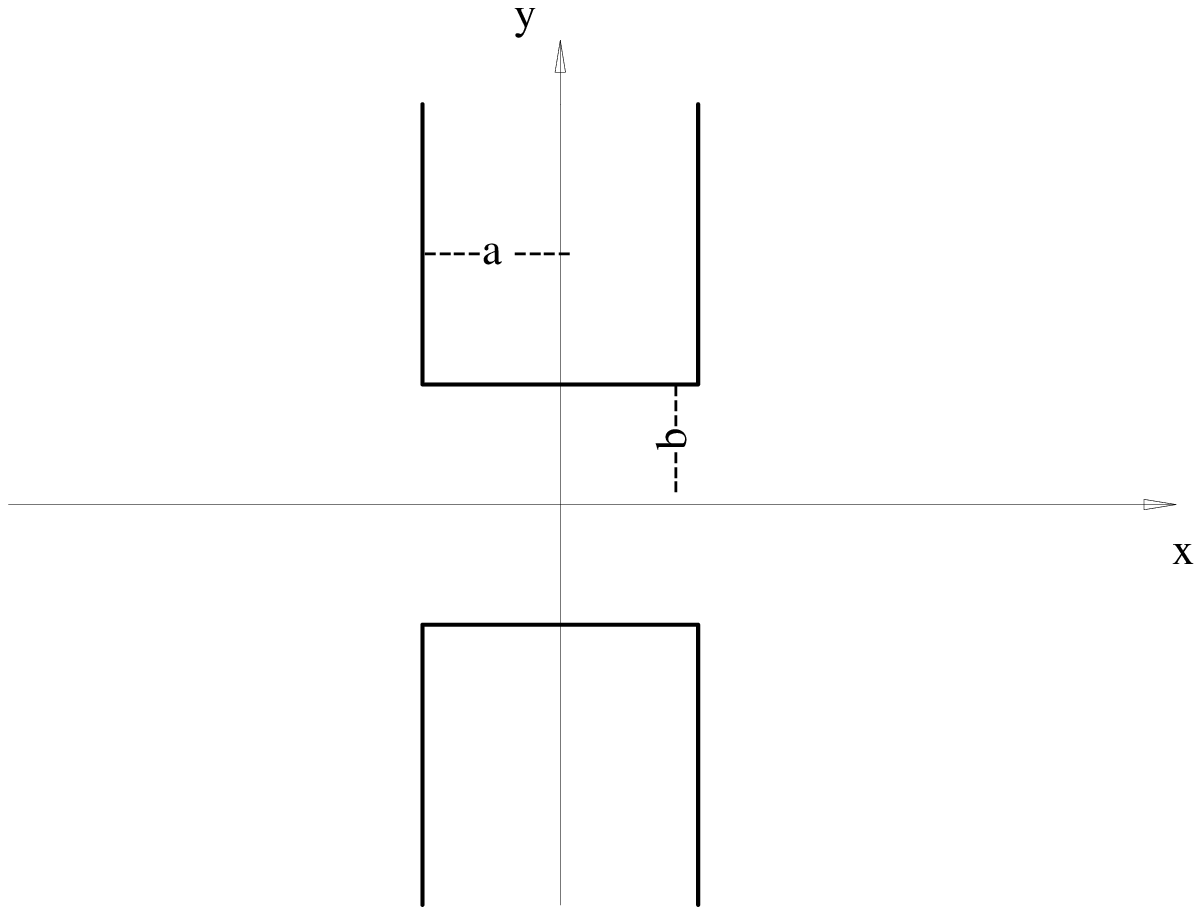}
\vsize=5 cm
\caption{\sl slit geometry, {\sl 2 a} is the length of the slit and {\sl
2 b} the width.}
\label{fig1}
\end{figure}

The scattering event commences at t=0 with a minimal uncertainty wavepacket

\bea\label{packet}
\Psi_0~&=&~A~e^w\nono
w&=&i~q_x~(x-x_0)+i~q_y~(y-y_0)-\frac{(x-x_0)^2}{4\s_1^2}-
\frac{(y-y_0)^2}{4~\s_2^2}
\eea

centered at a location $x_0,~y_0$ large enough for the packet to be
almost entirely outside the slit, 
$\s_1,~\s_2$ denote the width parameters of the packet
in the direction of the {\sl x, y} axes respectively, {\sl A } 
is a normalization constant, and, 
$ q_{x,y} = mv_{x,y}$ are the average momenta of the packet.

The algorithm for the numerical integration of the two-dimensional 
Schr\"odinger equation of the present work, 
is a direct extension and refinement of the methods of
refs. \ci{k1,k2,k3}.\footnote{
The price paid is an extremely lengthy calculation time of almost a day
of computer time in the VAX $\alpha$ cluster of the computation center
of the Hebrew University in Jerusalem for each and every run.}
The conservation of flux is verified by checking 
the normalization of the wave at the end of the process.
Typically, the flux is conserved to an accuracy better than $10^{-9}$ in
all the runs.

Figure 2 depicts a 3-D surface plot of the amplitude of the wave after
t=300 with parameters $x_0=-10,~y_0=0,~v_x=0.05,~v_y=0,~\s_1=1,
~\s_2=1$ scattering off a slit with dimensions {\sl a=2, b=3} wider than
the packet widths. The location of the slit is noted.
The repulsive potential strength of the screen is $V_{screen}=10^{19}$, 
effectively infinite, but still amenable to numerical treatment.

The hilly structure in the backward region is the wavepacket diffraction
in space and time studied in \ci{k1,k2,k3,k4}.

\begin{figure}
\epsffile{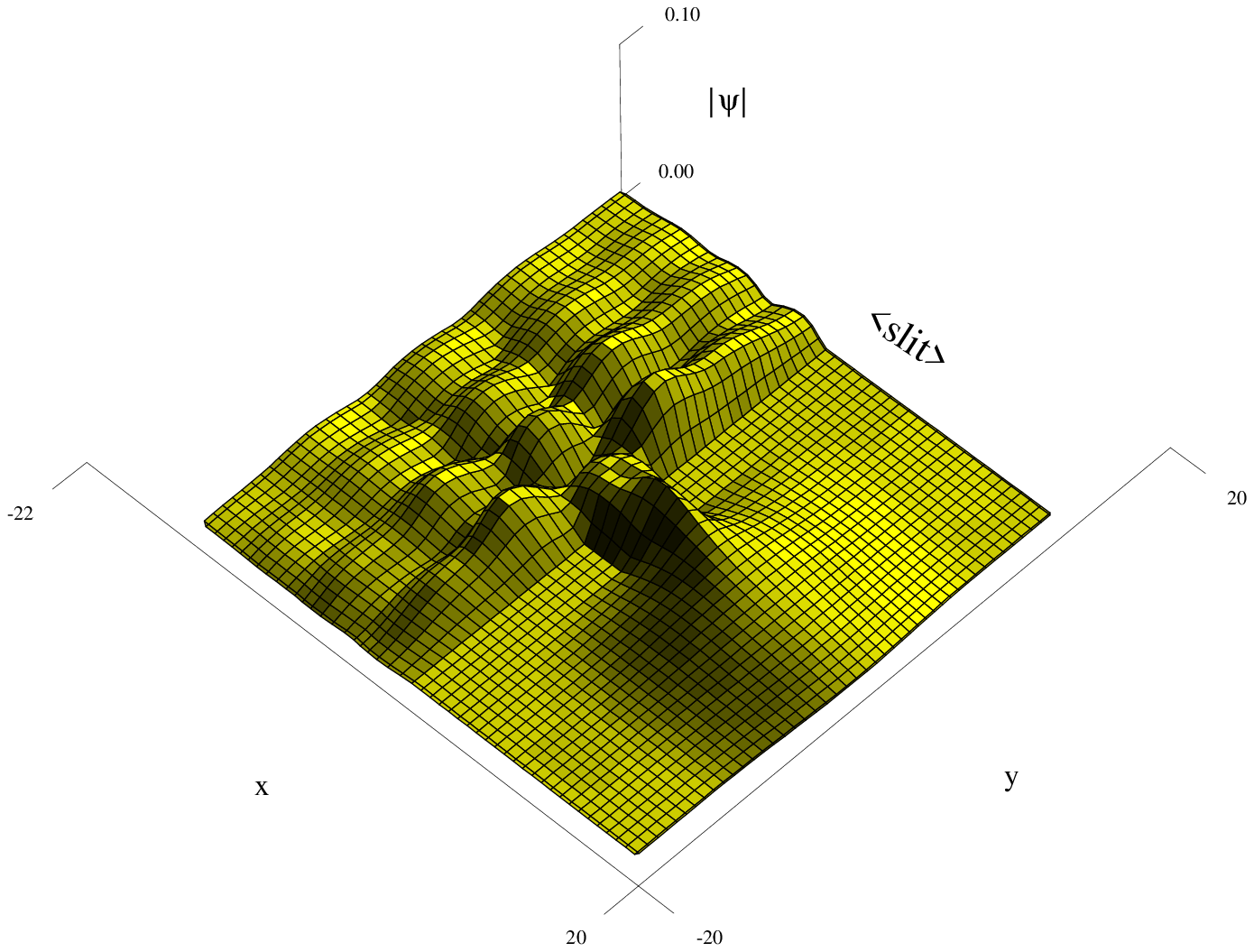}
\vsize=5 cm
\caption{\sl Single slit 3-D contour surface plot of the amplitude of
the wave , for a thin packet}
\label{fig2}
\end{figure}

The diffractive structure exists even
in a region facing the slit. This is in line with what found
before \ci{k3}, concerning the appearance of
wavepacket diffraction in space and time both for wells and barriers,
 as the opening may be considered to be a well inside a barrier.

In \ci{k1,k2,k3,k4}, it was found that for a wide packet, the
diffraction structure disappears. The same happens in the present 
case.
Figure 3 shows the scattering event for the same 
input as in figure 1, except for the
geometrical dimensions. The widths of the packet are now $\s_1=\s_2=2$,
 while the dimensions of the slit are $a=b=1$.

The hilly structure disappears, and, at the same time, the transmitted packet
is negligible as compared to the thinner packet.
Another feature that is evident from figures 2 and 3 is an interference pattern
along the {\sl y axis}. 

\begin{figure}
\epsffile{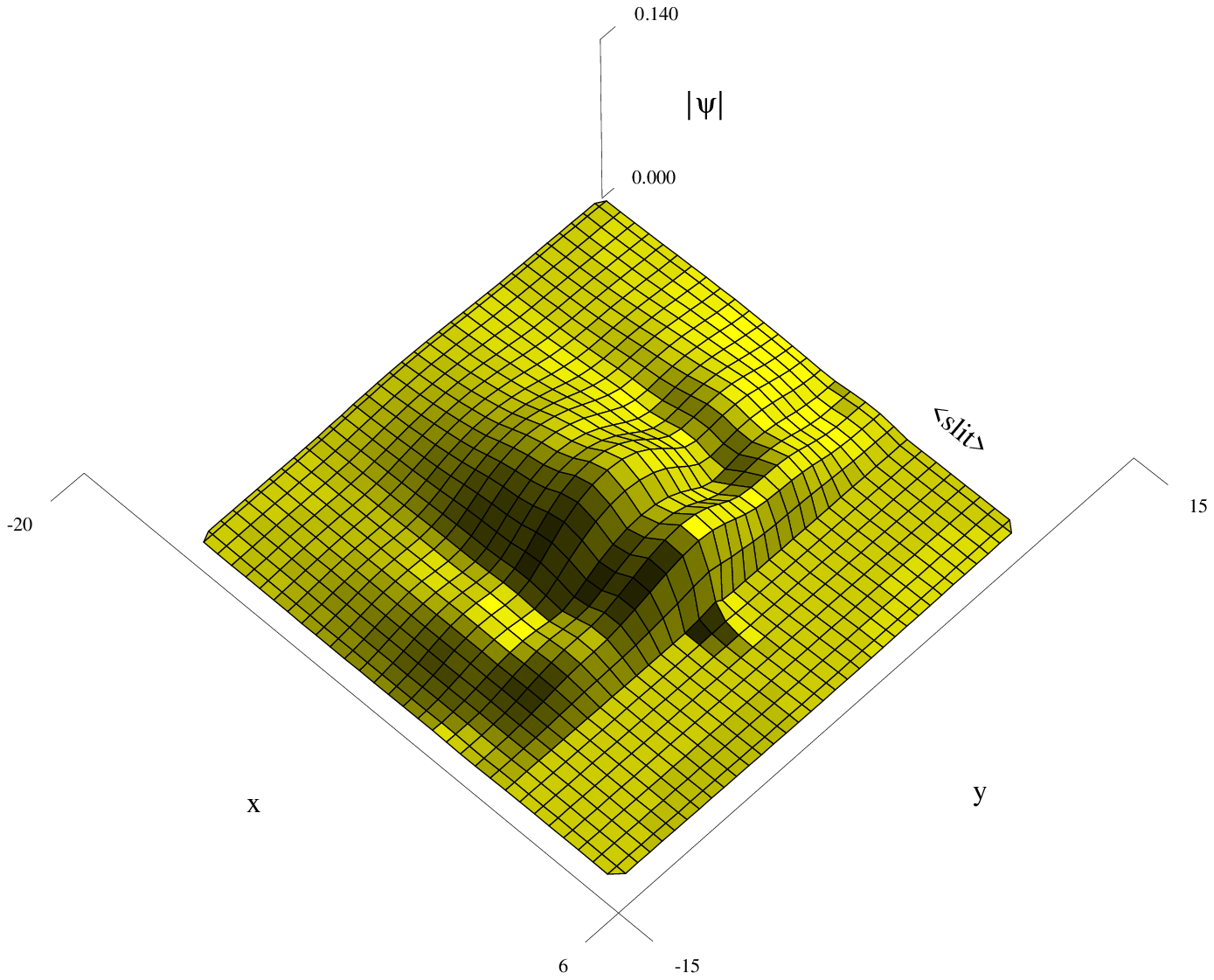}
\vsize=5 cm
\caption{\sl Single slit 3-D contour surface plot of the amplitude of
the wave, for a wide packet}
\label{fig3}
\end{figure}

\begin{figure}
\epsffile{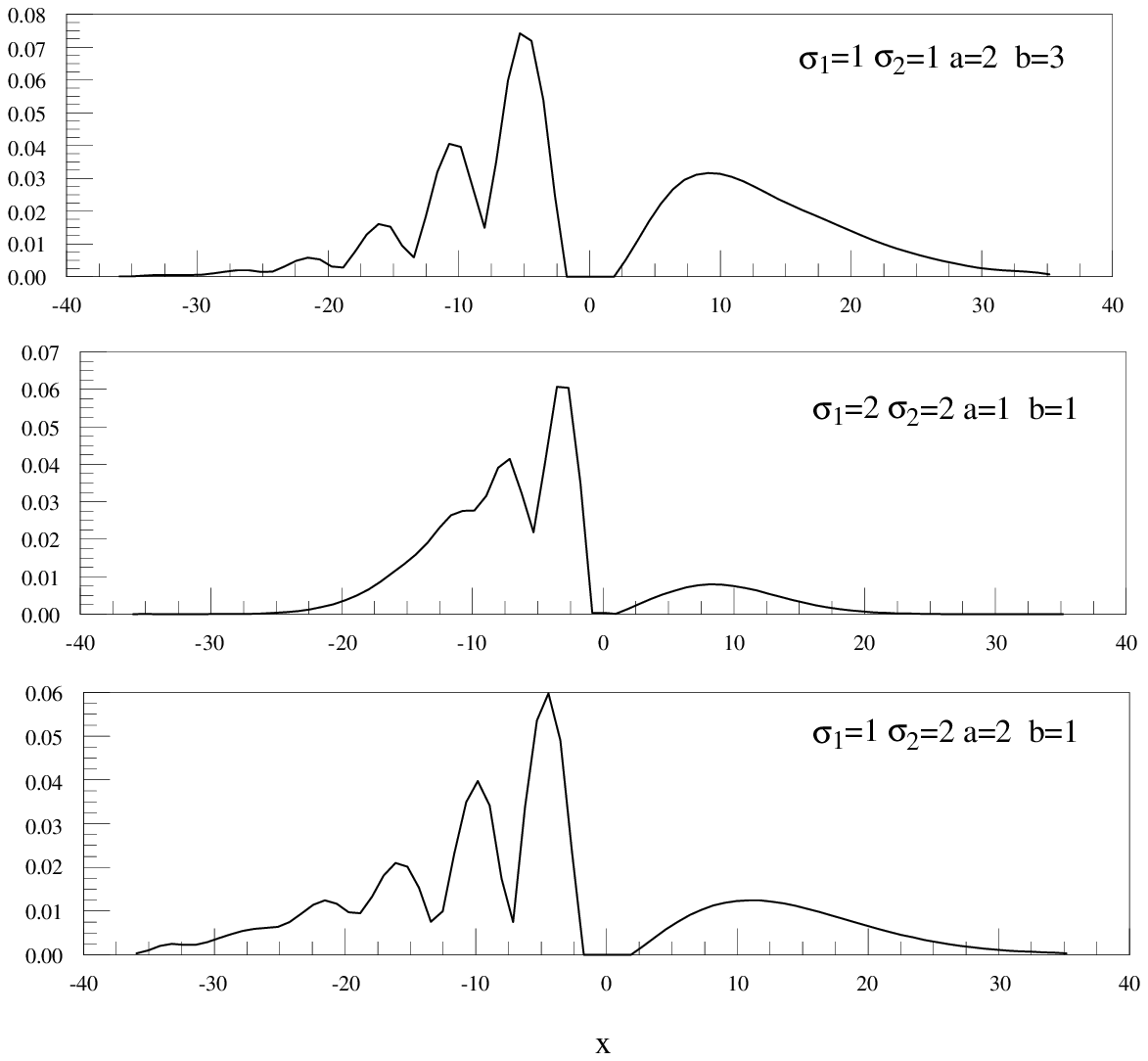}
\vsize=5 cm
\caption{\sl Wave amplitude as a function of {\sl x}
at a fixed {\sl y = 5.46} for various width and slit parameters}
\label{fig4}
\end{figure}

The disappearance of the multiple peak structure is gradual and is
influenced by the widths of the packet in the directions of both
axes. The relevant parameter that determines the diffraction structure
in the backward direction is the width of the packet in the
horizontal direction.
Even for relatively wide packets along the vertical axis, 
the diffraction pattern still exits.
This is evidenced in figure 4 where the same scattering event as figure 1
is shown for different packet width parameters and slit dimensions.

\vspace{2 cm}
Consider now a Young double slit arrangement as in figure 5.

\begin{figure}
\epsffile{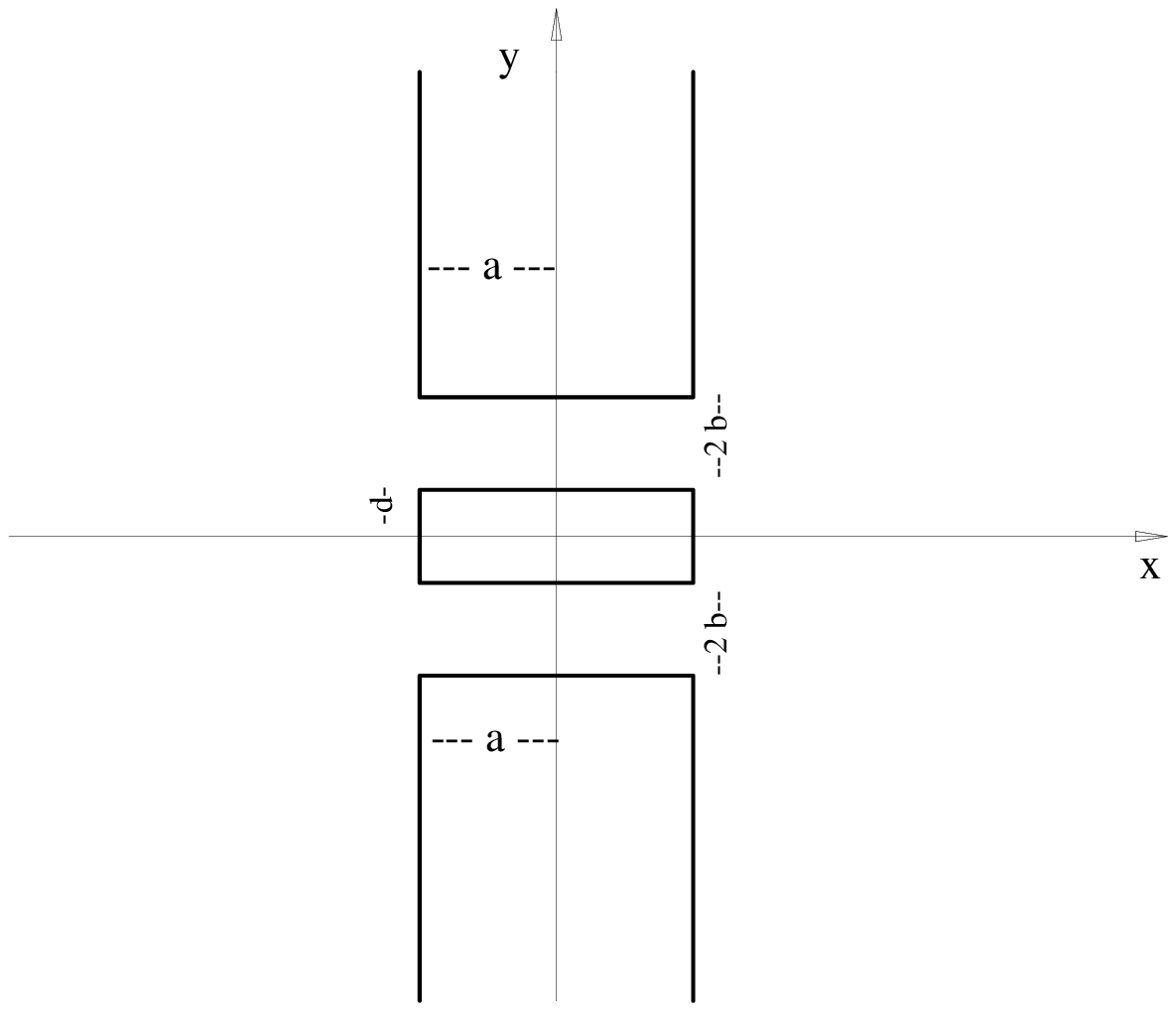}
\vsize=5 cm
\caption{\sl Geometry of the double slit arrangement}
\label{fig5}
\end{figure}

Two scattering cases are depicted in figure 6 and figure 7.
Figure 6 corresponds to the geometrical parameters for the
slit a = b = d = 2, and the packet widths $\s_1~=~\s_2~=1$.
Figure 7 corresponds to the geometrical parameters for the
slit a = b = 2, d = 2 and the widths $\s_1~=4~\s_2~=5$.

\begin{figure}
\epsffile{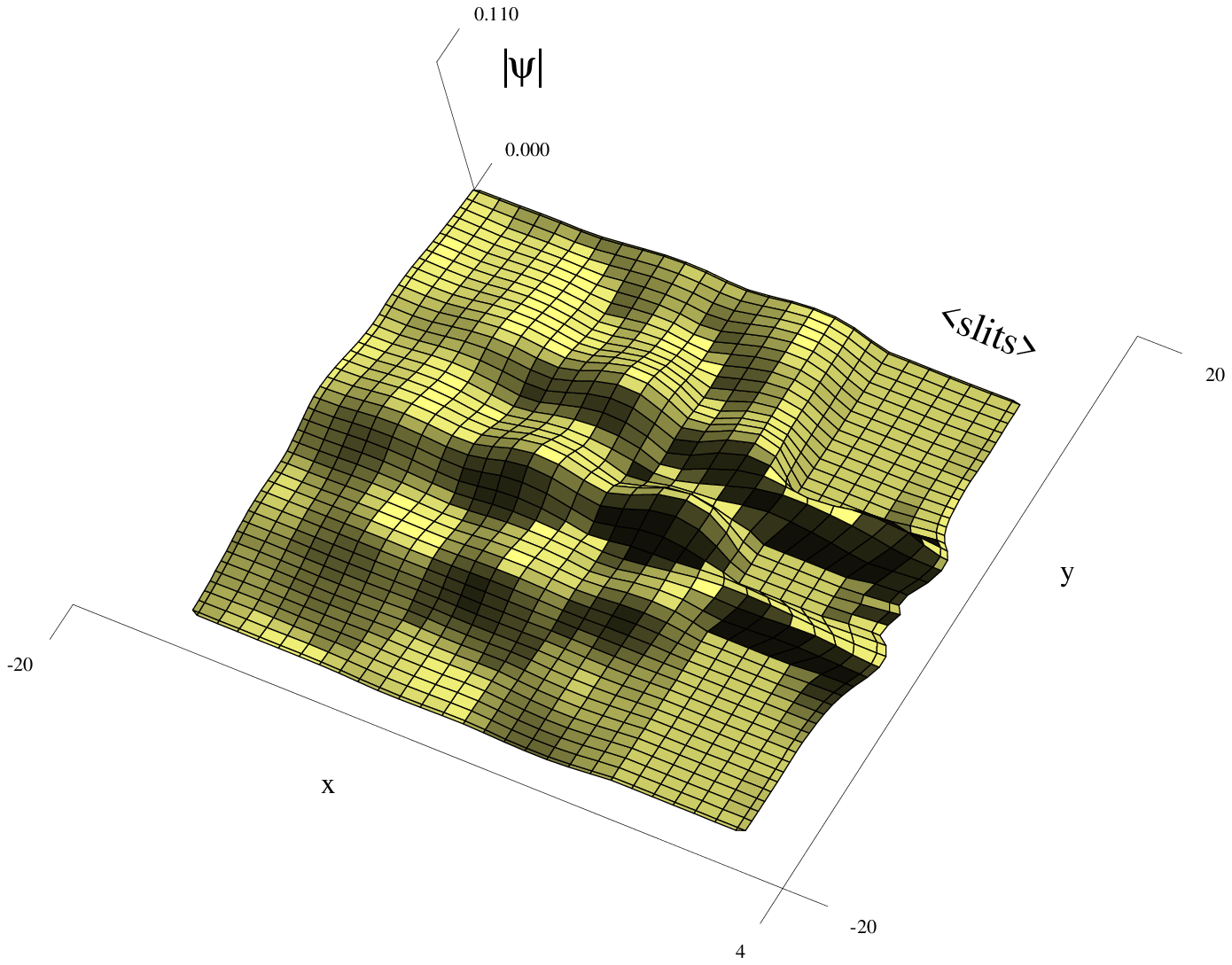}
\vsize=5 cm
\caption{\sl Double slit 3-D contour surface plot of the amplitude of
the wave in the backward zone, for a thin packet}
\label{fig6}

\end{figure}
\begin{figure}
\epsffile{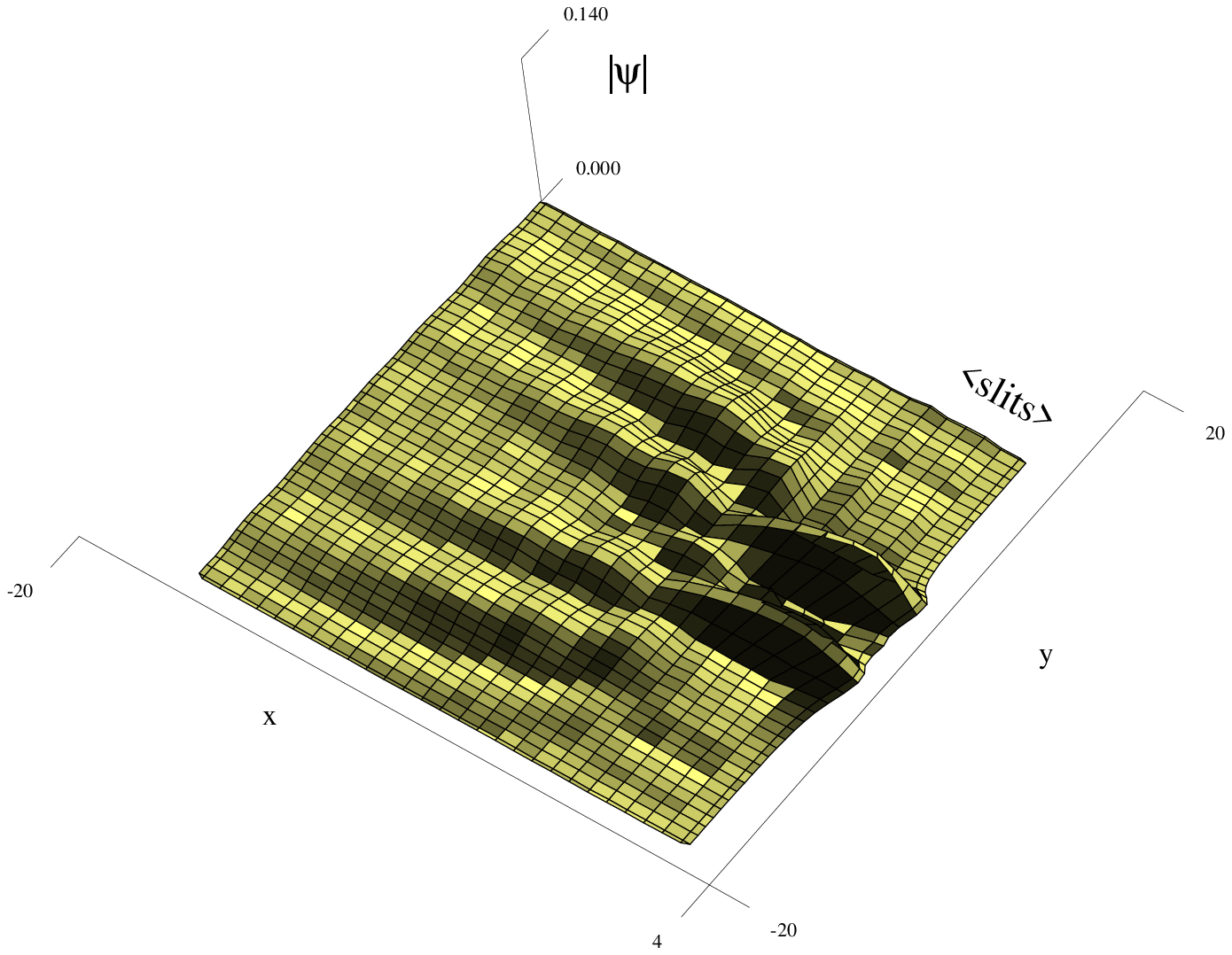}
\vsize=5 cm
\caption{\sl Double slit 3-D contour surface plot of the amplitude of
the wave in the backward zone, for a wide packet}
\label{fig7}
\end{figure}

The polychotomous peak structure appears here also 
for thin packets only.

In summary, the phenomenon of
wavepacket diffraction in space and time applies to the
single and double slit arrangements. The crucial
parameter for its appearance is the thickness of the slit and screen.
There exists also a diffraction pattern along
the axis parallel to the screen. The origin of this pattern is 
evidently due to the interference between incoming and reflected
waves, as well as the waves excited inside the slit cavity.
At this stage it was not possible to reproduce the qualitative  
features of this backward transverse pattern.

\section{\sl Single slit and double slit diffraction in the forward region}

In the forward direction, 
a very broad packet should behaves effectively, as a plane wave.
A broad packet must then produce a diffraction pattern.
On the other hand, 
very thin packets, should not generate a diffraction structure. 
Only a broad central peak is to be present. 
The other maxima of the diffraction pattern, have to 
disappear into the continuum because of destructive interference.
These assertions find support in the Kirchhoff
approximation to slit diffraction.

From Green's theorem, the Kirchhoff integral,
for the diffraction pattern from a slit 
with Dirichlet boundary conditions \ci{zei2}, may be readily written
down.
The wave function at a distance {\sl x, y} from the center of the slit
located at the origin of coordinates reads \ci{jackson,zei1}

\bea\label{green}
\Psi(\rr,t)&=&\frac{1}{4~\pi}\int_{0}^{t^+}~dt_0\int_{opening}d\vec{S_0}~
\cdot\bigg(
G(\rr,t,\rr_0,t_0)\bfnabla_0~\Psi(\rr_0,t_0)\nono
&-&\Psi(\rr_0,t_0)\bfnabla_0~G(\rr,t,\rr_0,t_0)\bigg)
\eea

where {\sl G} is the free propagator or Green's function for the
Schr\"odinger equation

\bea\label{propag}
G(\rr,t,\rr_0,t_0)&=&\Theta(\tau)~\sqrt{\frac{m}{2~\pi~i~\tau^3}}~
e^{\delta(\tau)}\nono
\delta(t)&=&{i~m~\frac{|\rr-\rr_0|^2}{2~\tau}}
\eea

$\Theta$, is the Heaviside step function that guarantees causality, {\sl m}
is the mass of the particle, $\tau=t-t_0$, 
with $\hbar=1$.
The $\bfnabla$ operator acts in the direction
perpendicular to the opening, the {\sl x} axis in the present case. 
The prediction of the diffraction pattern reduces now 
to the knowledge of the wave at the slit.

The next step, that is usually taken at this stage , is referred to, as the 
{\sl Kirchhoff approximation} \ci{jackson}. 
The Kirchhoff approximation uses for the wave at the slit, the
same expression as the incoming wave behind the slit.
This approach yields satisfactory results for plane waves and captures
the main features of the diffraction patterns.
In some cases, it fails, and a full solution of the problem is needed. 
The Green's function approach becomes then of a limited value, similarly
to the limitations of the Born approximation in potential scattering.
For the Schr\"odinger equation the results for plane wave diffraction from
slits using the Kirchhoff approximation \ci{zei1,zei2}, appear to be quite 
reasonable. It is then appropriate, to take advantage of the above equation 
(\ref{green}) for the case of wavepackets also.

The evaluation of the integrals of equation(\ref{green}), is
performed by resorting to
the stratagem of rotating the time axis $t_0$. An
alternative, would be to integrate it using Fresnel integrals in the far
field region. However, the interest of the present work is to compare
 to numerical results at all all times and positions. 

The rotation of the time axis, should avoid crossing of poles 
or even passing through the vicinity of poles, if no
additional compensating terms are desired. 
The direction of the rotation has to be such that, 
the integral still converges.
This technique permits to 
extend the integration over $t_0$, up until exactly $t_0~=~t$. 
The method was tested against known diffraction
expressions and found to be very accurate. Typically, a rotation by
an angle of $\phi(t_0)\approx0.001$, with around 10000 integration points
along the $t_0$ axis using double precision complex variables was used.

For the slit of figure 1 that extends to infinity in the {\sl z} direction,
and the packet of eq.(\ref{packet}), at $t_0\ne0$, the wave function becomes

\bea\label{kirch1}
\Psi(\rr,t)&=&\frac{-i~m~x}{4~\pi}~
\int_{0}^{t^+}~dt_0\int_{-b}^{b}~dy_0~\Psi(0,y_0,t_0)
\frac{e^{\delta(\tau)}}{\tau^2}~\bigg(1-~\frac{i\tau~(x_{00}+v~t_0)}
{2~m~x~({\s_1}^2+\frac{i~t_0}{2~m}})\bigg)\nono
\delta(\tau)&=&i~m~\frac{(y-y_0)^2+x^2}{2~\tau}
\eea

with the incoming packet given by

\bea\label{packet1}
\Psi(0,y_0,t_0)&=&\sqrt{\frac{\s_1~\s_2}{2~\pi}}~\frac{e^{-u}}{d_1~d_2}\nono
u&=&\frac{(x_{00}+~v_x~t_0)^2}{4{\s_1}^2~d_1}+\frac{(y_0-y_{00}-v_y~t_0)^2}
{4~{\s_2}^2~d_2}\nono
d_1&=&{\s_1}^2+\frac{i~t_0}{2~m}\nono
d_2&=&{\s_2}^2+\frac{i~t_0}{2~m}
\eea

where now, $x_{00}, y_{00}$, denote the initial center of the packet.

Figure 8 shows the long-time 
diffraction patterns for a wide packet and a thin packet
in the Kirchhoff approximation of eq.(\ref{kirch1})

The wide packet $\s_1~=\s_2~=100,~x_0~=~-200$, 
indeed has a diffractive structure, 
the thin packet $\s_1~=\s_2~=0.5,~x_0~=~-10$, lacks it. 
The waves are normalized to 
one at {\sl y=0} and the slit dimensions are a=2, b=3.
The velocities are both taken to be v = 0.5, the mass is  m=20,
and the time is t=10000 at x=5000 as a function of {\sl y}.

\begin{figure}
\epsffile{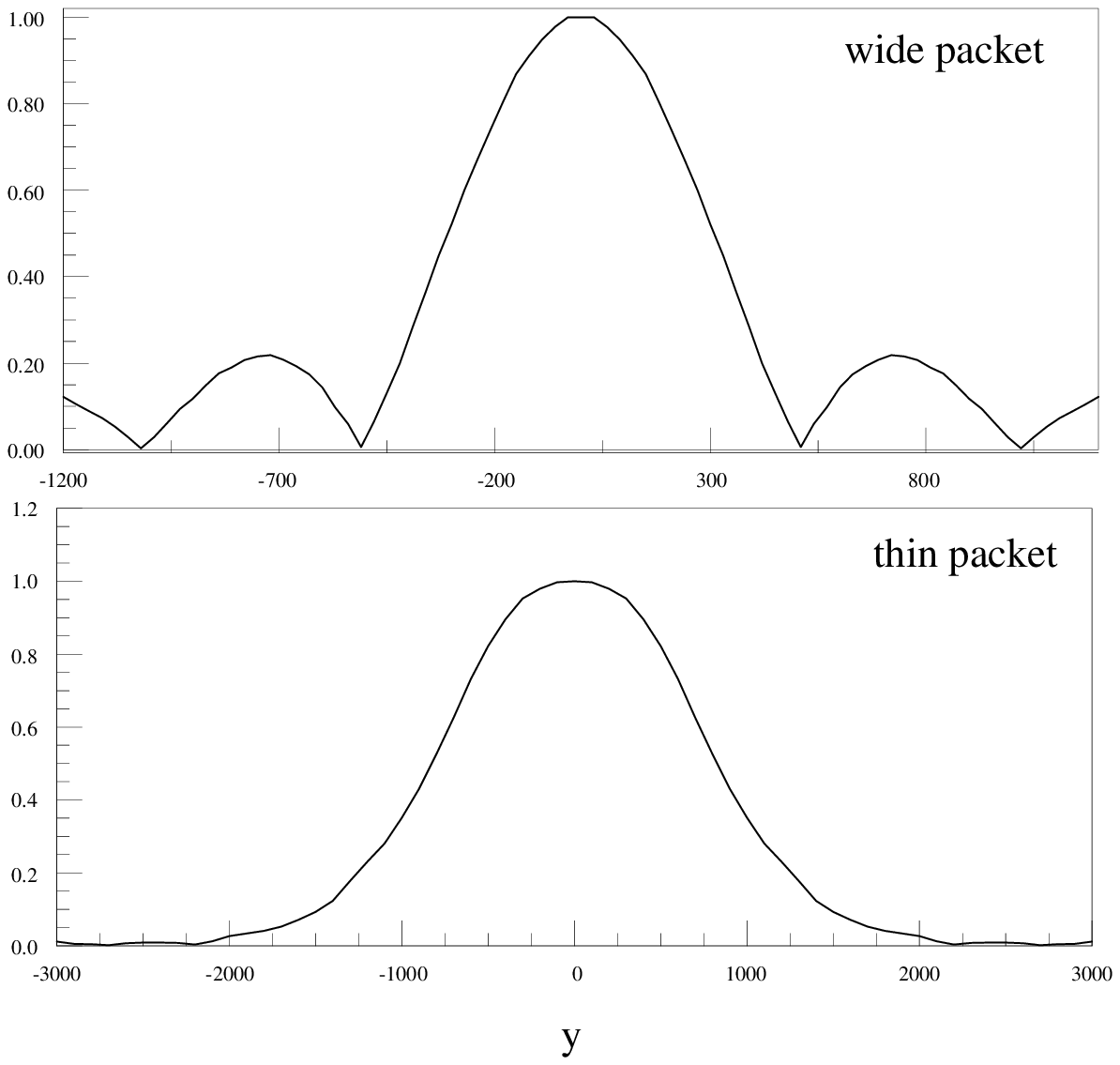}
\vsize=5 cm
\caption{\sl Kirchhoff approximation to the diffracted wave}
\label{fig8}
\end{figure}

The Kirchhoff approximation, considering
only the slit opening and not its width, generates the expected 
diffraction pattern for a wide packet and a broad peak for a thin one.

However, the numerically calculated waves at finite times do not
agree with the predictions of figure 8.
Figure 9 depicts the numerical results for the 
amplitude of the wave for a thin packet
at {\sl x=29.76} as a function of {\sl y} for the
scattering event of a single slit, at
t=300 with initial parameters $x_0=-10,~y_0=0,~v_x=0.05,~v_y=0.,~\s_1=1,
~\s_2=1$ scattering off a slit with dimensions {\sl a = 2, b = 3} wider than
the packet widths. The wave is renormalized to amplitude equal unity 
at {\sl y=0}.

For the thin packet, 
a broad peak was expected, but, instead, a structure resembling a
 diffraction pattern is obtained.

\begin{figure}
\epsffile{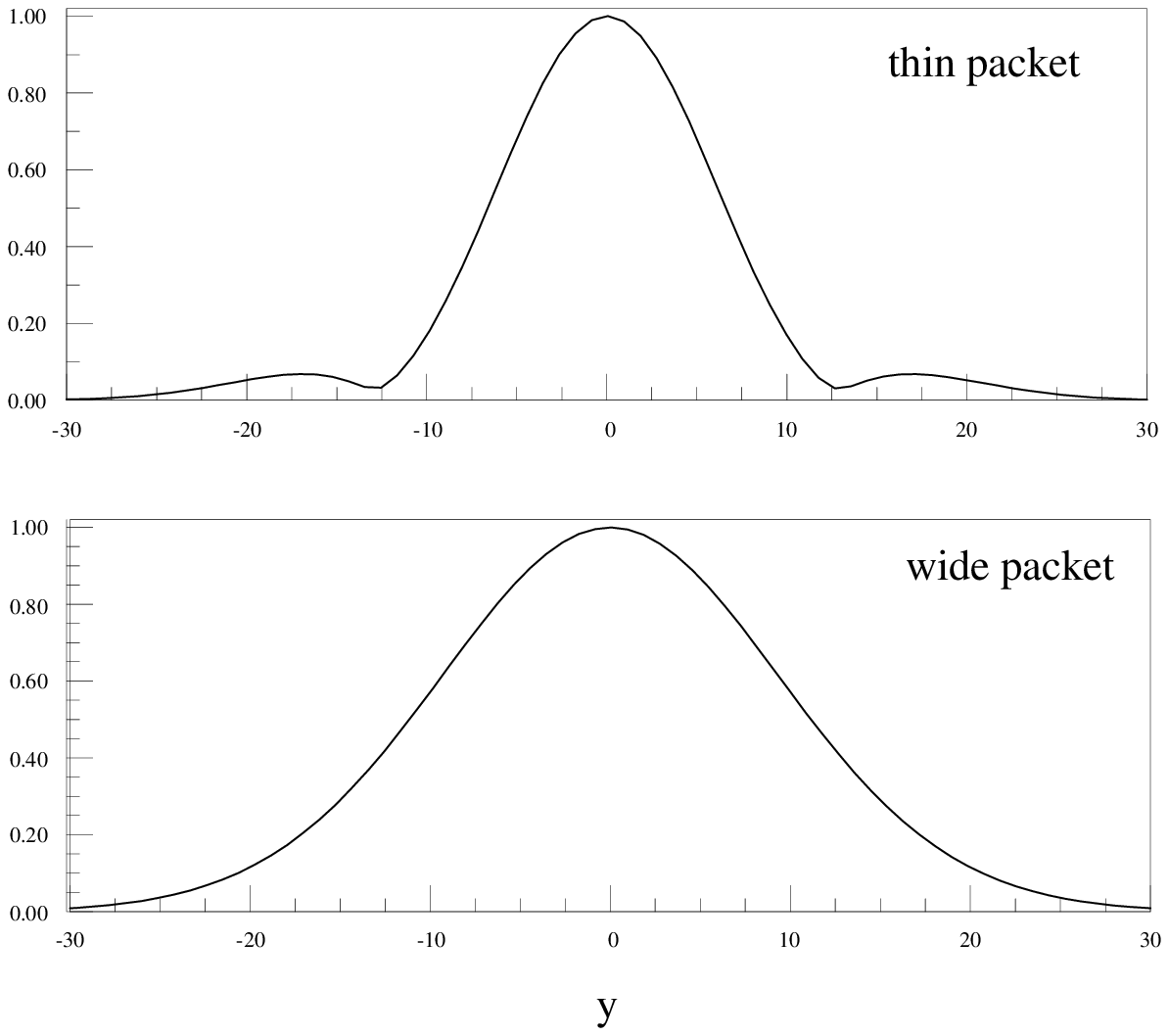}
\vsize=5 cm
\caption{\sl Numerical wave amplitude of the scattered wave at x= 29.76,
as a function of y, for a thin packet as compared to a broader packet}
\label{fig9}
\end{figure}

At the bottom of the figure, the diffraction pattern disappears
when the dimensions of the packet are large (but not infinite)
as compared to the slit. 
This graph corresponds to a packet and slit with
dimensions $\s_1~=~2~,\s_2~=~1$, a = 1, b = 2.

The numerical results for various input parameters, 
show that the crucial physical
parameter for the appearance of a diffraction pattern is the ratio 
$\frac{\s_1}{a}$. For ratios larger than one, no
pattern was observed. This is analogous to the behavior in the backward
region, for which a multiple peak structure exists and persists only for
packets that are initially thinner than the length of the slit.

The ratio $\frac{\s_2}{b}$, is also important, but the
sensitivity is much less pronounced. It has to be quite large
for the pattern to disappear.
Figure 10 exemplifies this property. The upper curve is the same as the upper
curve of figure 9, while the lower curve corresponds to a wide packet
in the {\sl y} direction, $\s_2~=5,~b~=1$. For $\s_2<5$, the diffraction 
pattern persists.

\begin{figure}
\epsffile{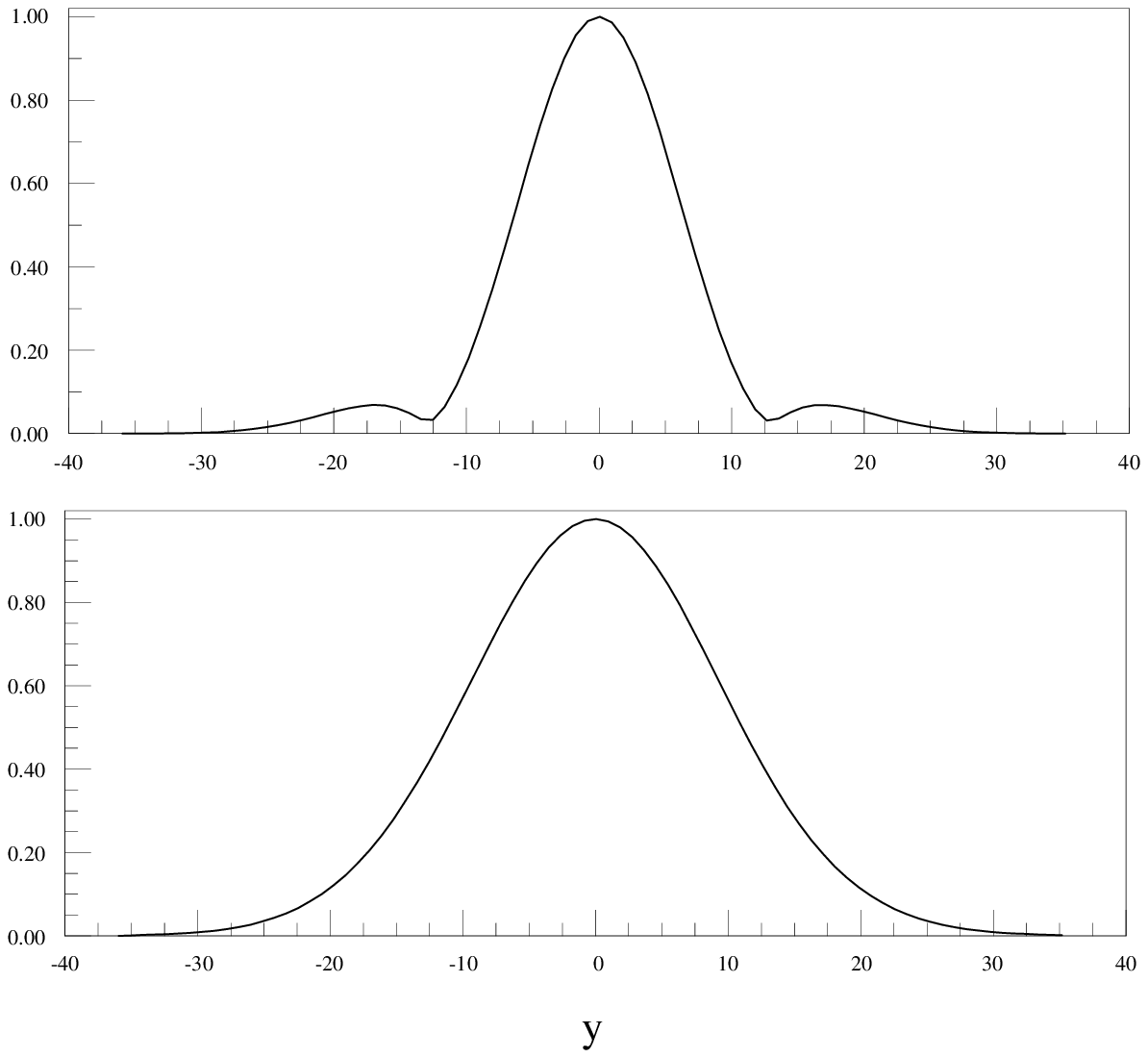}
\vsize=5 cm
\caption{\sl Numerical wave amplitude of the scattered wave at x= 29.76,
as a function of y, for a thin packet as compared to a broader packet
in the y direction}
\label{fig10}
\end{figure}

The forward diffraction along the vertical direction, 
appears to be connected to the phenomenon of
diffraction of wavepackets in space and time along the horizontal direction.
in the backward region.

The numerical results depicted in figures 9 and 10
suggest that, the intuitive assumption 
of no diffraction for a thin packet , is not
borne out by the Schr\"odinger equation evolution of the packet.
At the other end of plane waves, namely thin packets, somehow, 
diffraction reemerges. 

In figure 11 the Kirchhoff approximation of eq.(\ref{kirch1}) is calculated
for both the thin and broad packets.

\begin{figure}
\epsffile{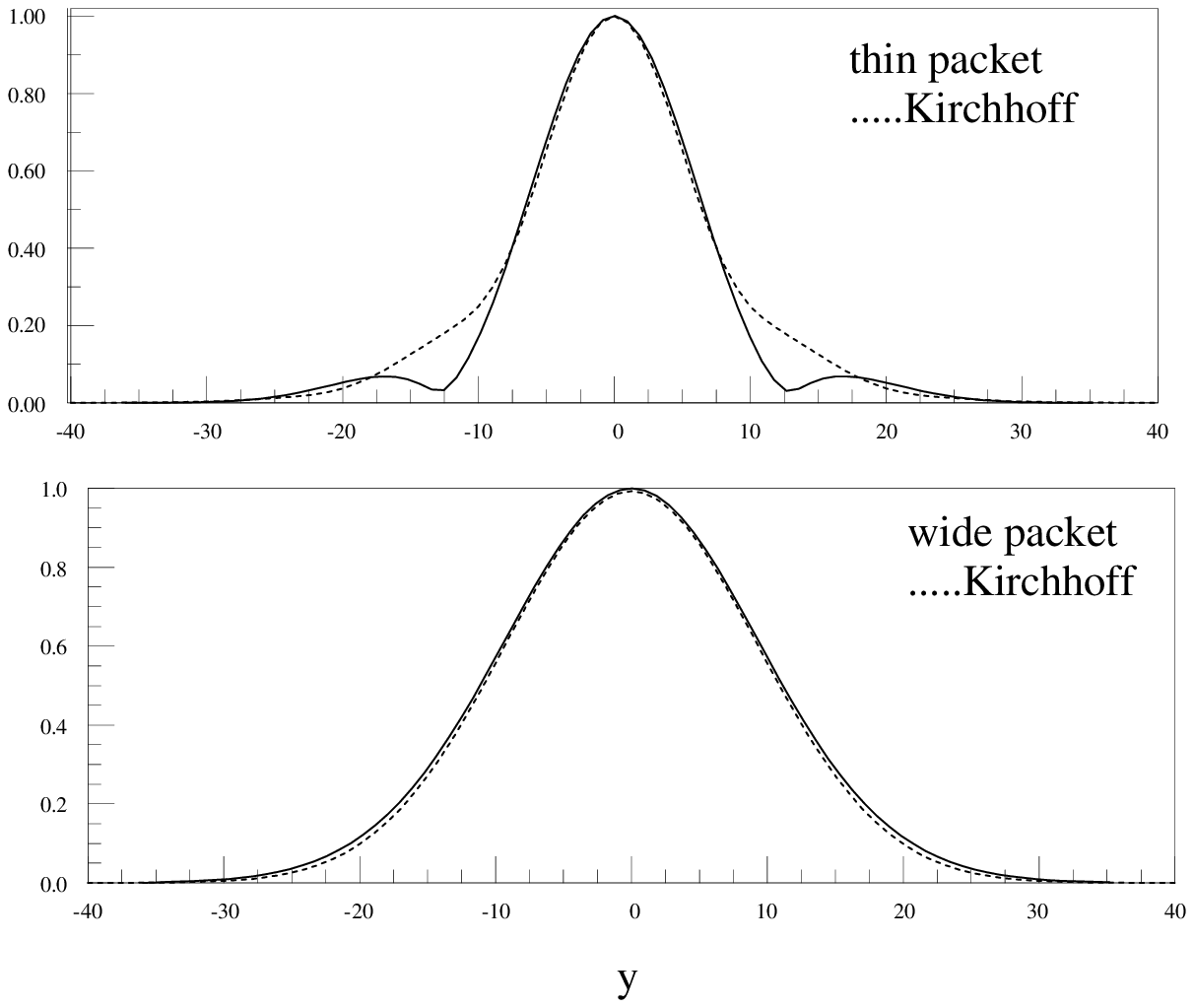}
\vsize=5 cm
\caption{\sl Comparison of the waves of figure 9 to the Kirchhoff
approximation}
\label{fig11}
\end{figure}

Although the shape is correct, the details of the diffraction structure
for a thin packet are not reproduced by the Kirchhoff approximation.
Many other numerical runs give the same results. The divide between
thin and broad packets makes itself evident in the appearance, or lack of, 
a forward diffraction pattern.
In the next section an improvement upon the Kirchhoff approximation is
developed.

\vspace{2 cm}

The results for the forward diffraction patterns in the double slit
 geometry of figure 5 are also very sensitive to the
initial widths of the packets.

Figure 12 depicts the wave amplitude at {\sl x = 35.16} as a function of
{\sl y}, for an impinging packet with parameters $\s_1~=~\s_2~=1, x_0~=~-10,~
y_0~=~0,~v~=~=0.05,~m~=~20$, at t = 300,
on the double slit of figure 5 with measures a = b = d =2.
The dotted line is the Kirchhoff approximation calculation of
equation (\ref{kirch1}), with appropriately
modified limits on the integration over $y_0$. 
The analytical results give qualitatively
the broad features of the interference and diffraction patterns, but
 as for the single slit, they miss the details.

\begin{figure}
\epsffile{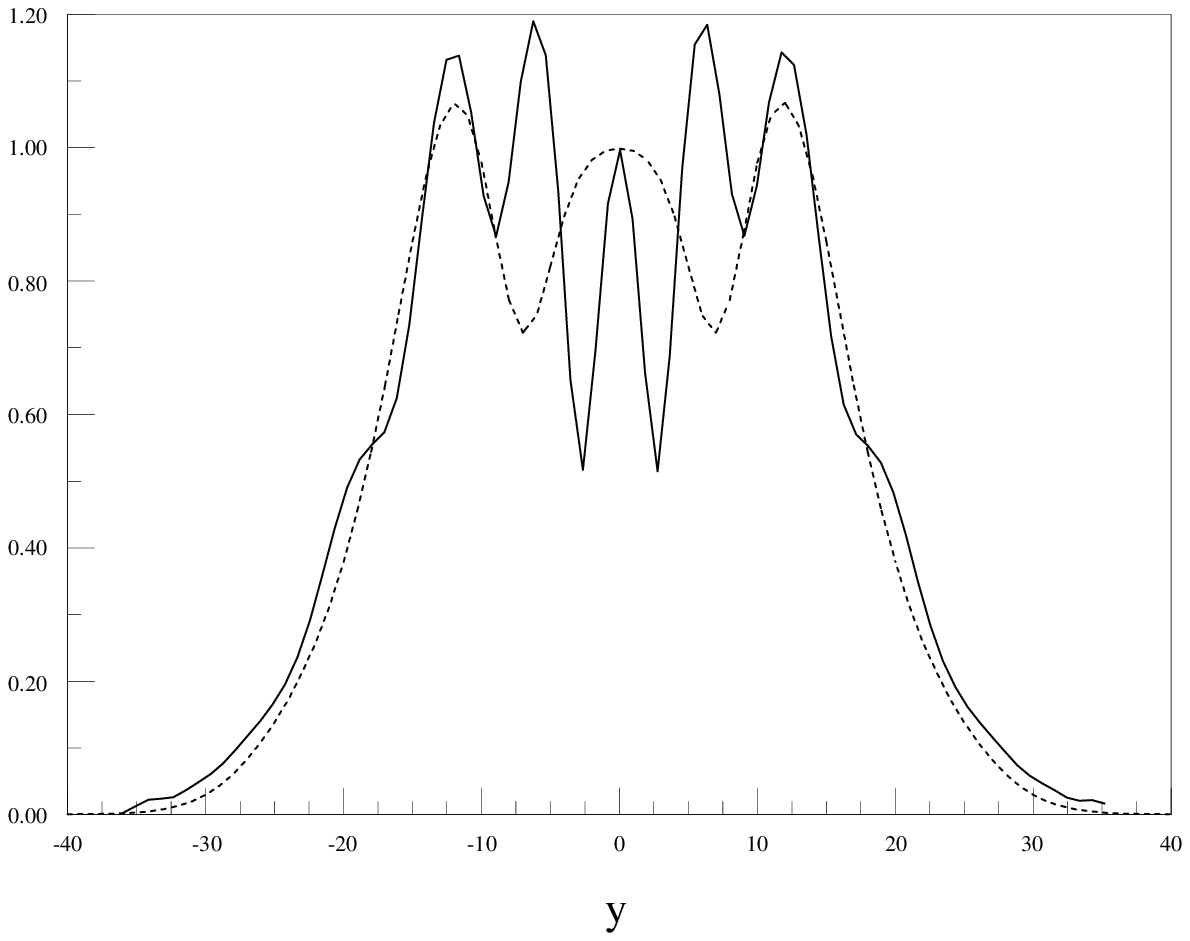}
\vsize=5 cm
\caption{\sl Double slit interference pattern at
{\sl x=35.16} as a function of {\sl y} for a thin packet, full line.
Kirchhoff approximation of eq.(\ref{kirch1}), dotted line}
\label{fig12}
\end{figure}

Figure 13 depicts the results for a wider packet on a smaller slit.
The parameters are here $\s_1~=4~\s_2~=5, x_0~=~-10,~
y_0~=~0,~v~=~=0.05,~m~=~20$, at t = 300, x~=~15
on the double slit of figure 5 with measures a = b = 2, d =1.
Reflections from
the 'side walls' of the enclosure that define the integration region produce
the spurious oscillations in the numerical results.
\footnote{Beyond {\sl x=15},
the oscillations increase and blurr the picture. For a more reliable
 numerical calculation, a much larger area of integration is needed.}

\begin{figure}
\epsffile{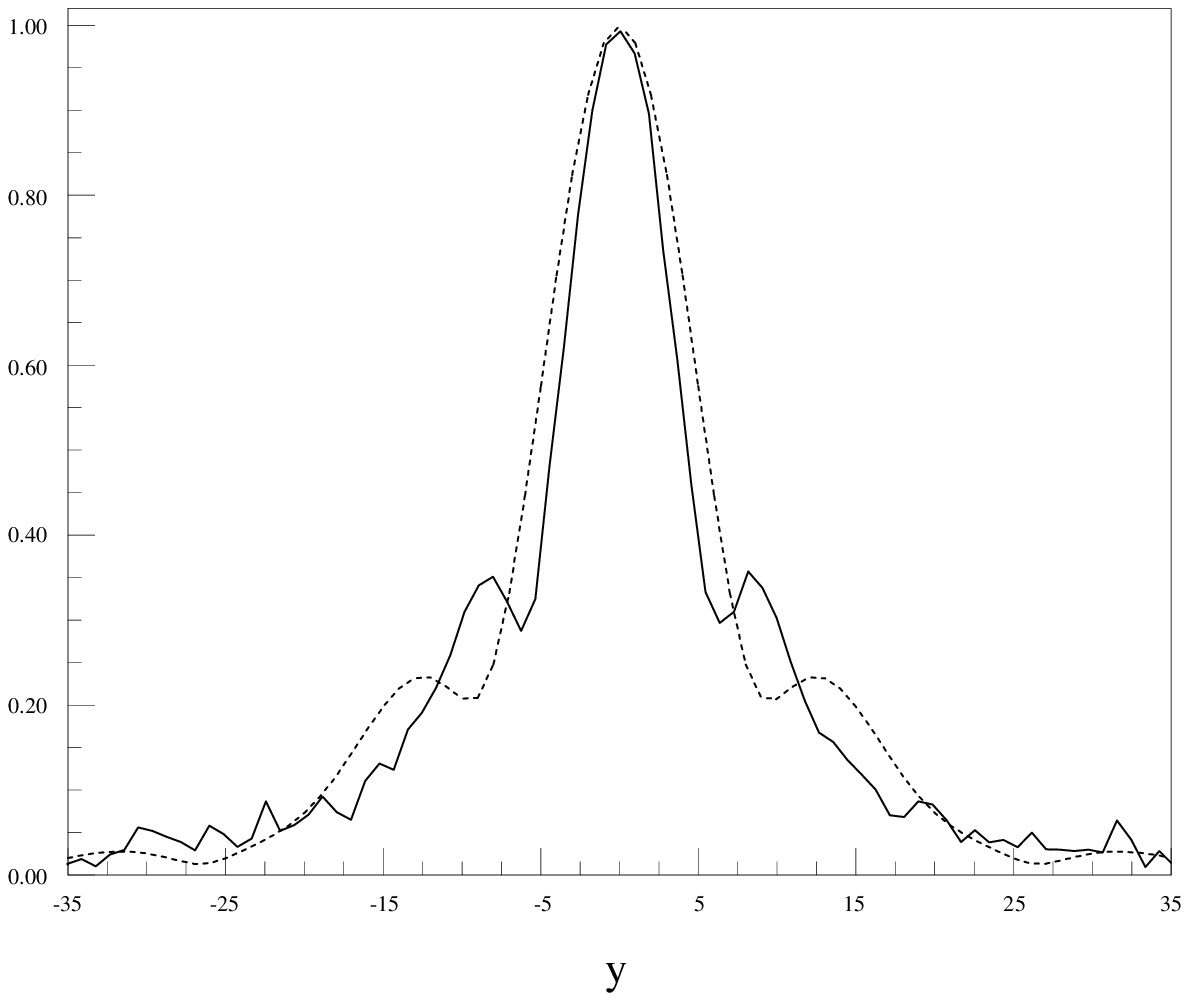}
\vsize=5 cm
\caption{\sl Double slit interference pattern at
{\sl x=15} as a function of {\sl y} for a wide packet, full line.
Kirchhoff approximation of eq.(\ref{kirch1}), dotted line}
\label{fig13}
\end{figure}

The dotted curve corresponds to the Kirchhoff approximation eq.(\ref{kirch1}).
Contrarily to the thin packet case, the 
Kirchhoff approximation captures the main features of the pattern.

Figures 12 and 13, are consistent with the results for 
a single slit. The dimensions of the packets are relevant to the
diffraction and interference structures seen. Thin
packets generate diffraction and interference structures that are absent
for wider packets. This effect is even more evident when 
the comparison to the Kirchhoff approximation is made. 
The Kirchhoff approximation works well
for wide packets and misses the numerical results for thin
packets. 
A very essential part of the
scattered wave at the slit, is not included in the Kirchhoff integral.

In section 4 it is proposed, that this element may be, 
the excitation of cavity modes inside the slit.

\section{Cavity modes diffraction}

The exact solutions to the single slit problem are unknown, even
for plane waves. It is intuitively apparent that the wavepacket
 inside the slit comprises transient waves and
standing waves, as an electromagnetic wave in a Fabry-Perot interferometer 
\ci{Born}.
The transient waves are already included in the Kirchhoff approximation.
However, the cavity modes standing waves, are not.
The continuous spectrum of the incoming packet is
decomposed inside the slit, into, a continuous background and 
a discrete spectrum. 
The reflected and transmitted waves take care 
of the continuity at the entrance and exit of the slit.
The amplitude of the cavity modes has to decay with time.
This behavior was seen in the numerical solutions of \ci{k1,k2}.

The cavity modes are already included in the spectrum
of the incoming wave. However, there, they do not have a prominent
weight as compared to any other mode.
Despite some amount of double counting that may be introduced
by adding a specific contribution that singles out the cavity modes,
and, in the absence of an alternative way,
the full wave at the slit will be taken to be the sum 
of the transient wave, the incoming packet with a continuum of
modes, and the standing waves with a discrete spectrum, the cavity modes. 
This procedure is somewhat
analogous to the expansion of the wave at the slit, 
in terms of the spectrum of a very deep well.
This well may readily be disclosed by writing the cavity potential, that is
zero, as the large
repulsive barrier of the screen, in addition to an attractive well of the same
value. 

The cavity modes with a Dirichlet boundary condition
on the walls and the condition of having a standing wave with an antinode at
the right end of the slit (without end correction) amount to 
the expansion

\bea\label{psicav}
\Psi_{cavity}&=&\sum_{p_n}~A_n~cos(p_n(x-a))~\sum_{q_n}~B_n~cos(q_n~y)\nono
p_n&=&\frac{n\pi}{2~a}\nono
q_n&=&\frac{(2~n+1)\pi}{2~b}
\eea

Where the origin has been shifted to the right hand side of the slit.
In order to find the coefficients, the {\sl unknown} solution
outside the slit-cavity is needed. 
At the left end of the cavity {\sl x = - a}, the wave consists of 
the incoming wave, the reflected wave and the transmitted wave, both,
the continuum parts and the cavity modes.
In a fully consistent treatment, both the function and the derivative
have to be continuous. However, for the implementation of both continuity
conditions, a respectable ansatz for the waves in all regions has to be
assumed. In the absence of such an ansatz, the continuity of the function, 
 or the derivative -but not both-, 
is able to determine, up to an overall scale, the $B_n$ coefficients. The
$A_n$ coefficients are undetermined. The latter were fixed by
 comparing to the  
weight factors of the Fourier decomposition of
the incoming wave.\footnote{Other functional choices, like
{\sl sin} function, or exponential, or mixtures, gave inconsistent results
for both a wide packet and a thin one.}

Up to an overall unknown phase (possibly time dependent), the amplitudes are 
then given by

\bea\label{coefs}
A_n~B_l&=&\sqrt{\frac{\s_1~\s_2}{2~\pi}}~e^{-u}
~cos(p_n(x_0-a))\nono
u&=&-\s_1^2~(p_n-m~v)^2-\s_2^2~q_l^2-i\frac{p_n^2+q_l^2}{2~m}~t
\eea

The total wave function is then $\Psi_{total}
~=~\Psi_{incoming}~+~C~\Psi_{cavity}$.
Where $\Psi_{incoming}$ at the right hand end of the slit is given in
eq.(\ref{kirch1}), and {\sl C} is an unknown complex number. 
Comparing to the Fourier expansion of the incoming wave,
 the absolute value of this number should be of the order of 
$|C|~\approx~\frac{\pi^2}{4~a~b}$, the product of 
wavenumber increments.

Fortunately, the cavity modes and continuum contributions 
affect different regimes. The cavity mode piece is
dominant for $\s_1\approx\s_2\approx0$, the continuum 
contribution is negligible there, Eqs.(\ref{packet},\ref{coefs}).
It is then not too worrisome that, the exact value of {\sl C}
is not known. 
$\Psi_{total}$ is now introduced in eq.(\ref{green}) in the slot
of the wave function at the opening, the rest of the calculation proceeds
as for the case of the Kirchhoff approximation explained above.

Figure 14 shows the numerical results together with $\Psi_{total}$,
for the single slit diffraction pattern
previously depicted in figures 9 and 11. It was shown there, that the Kirchhoff
approximation misses the diffraction pattern for a thin packet, whereas
it reproduces the numerical calculation quite accurately for a wide one.

From figure 14 it is clear that, the cavity modes play an important role
in reproducing the thin packet results, while, at the same time, they
do not spoil the fit to the wide packet case, as expected.

The cavity modes are then, an important ingredient in the
diffraction pattern of thin packets, and are almost negligible for
a wide packet. 

\begin{figure}
\epsffile{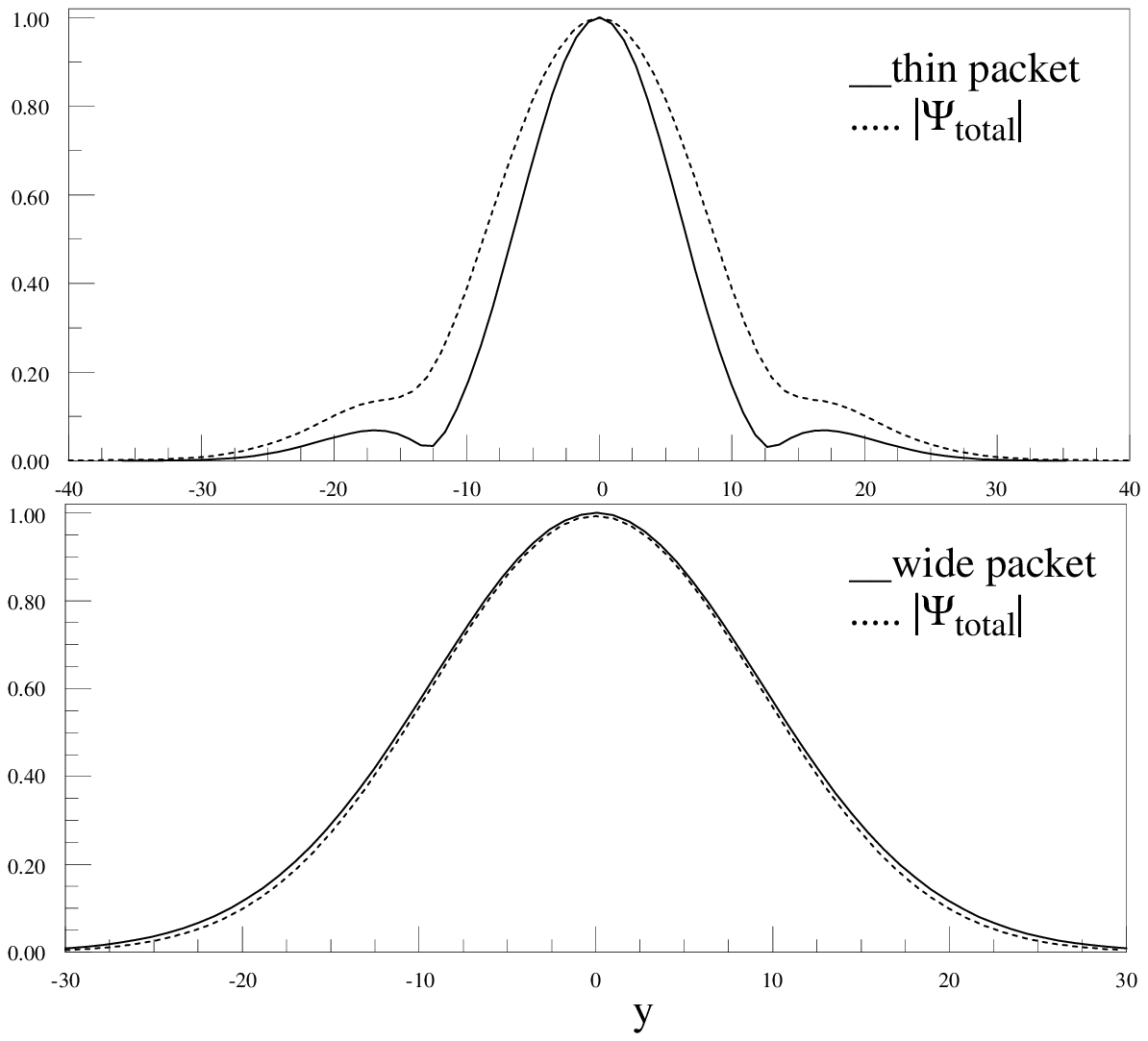}
\vsize=5 cm
\caption{\sl Comparison of the waves of figure 9 to the cavity mode
plus incoming approximation, $\Psi_{total}$}
\label{fig14}
\end{figure}

After gaining some confidence with the approach, it is possible to look at
the long time behavior of the diffraction pattern. An aspect that cannot
be obtained by numerical treatment at the present time.
Although the results are presumably inaccurate quantitatively, the
 structure is expected to be relatively reliable.
The calculation is presented in figure 15.
The parameters are here $\s_1~=~\s_2~=1, x_0~=~-10,~
y_0~=~0,~v~=0.05,~m~=~20$, at t = 50000, x~=~5000,
on the single slit of figure 1 with measures a = 2, b = 3,
for the upper graph, and, $\s_1~=~\s_2~=100,~x_0~=~-200$ for the lower graph.
\begin{figure}
\epsffile{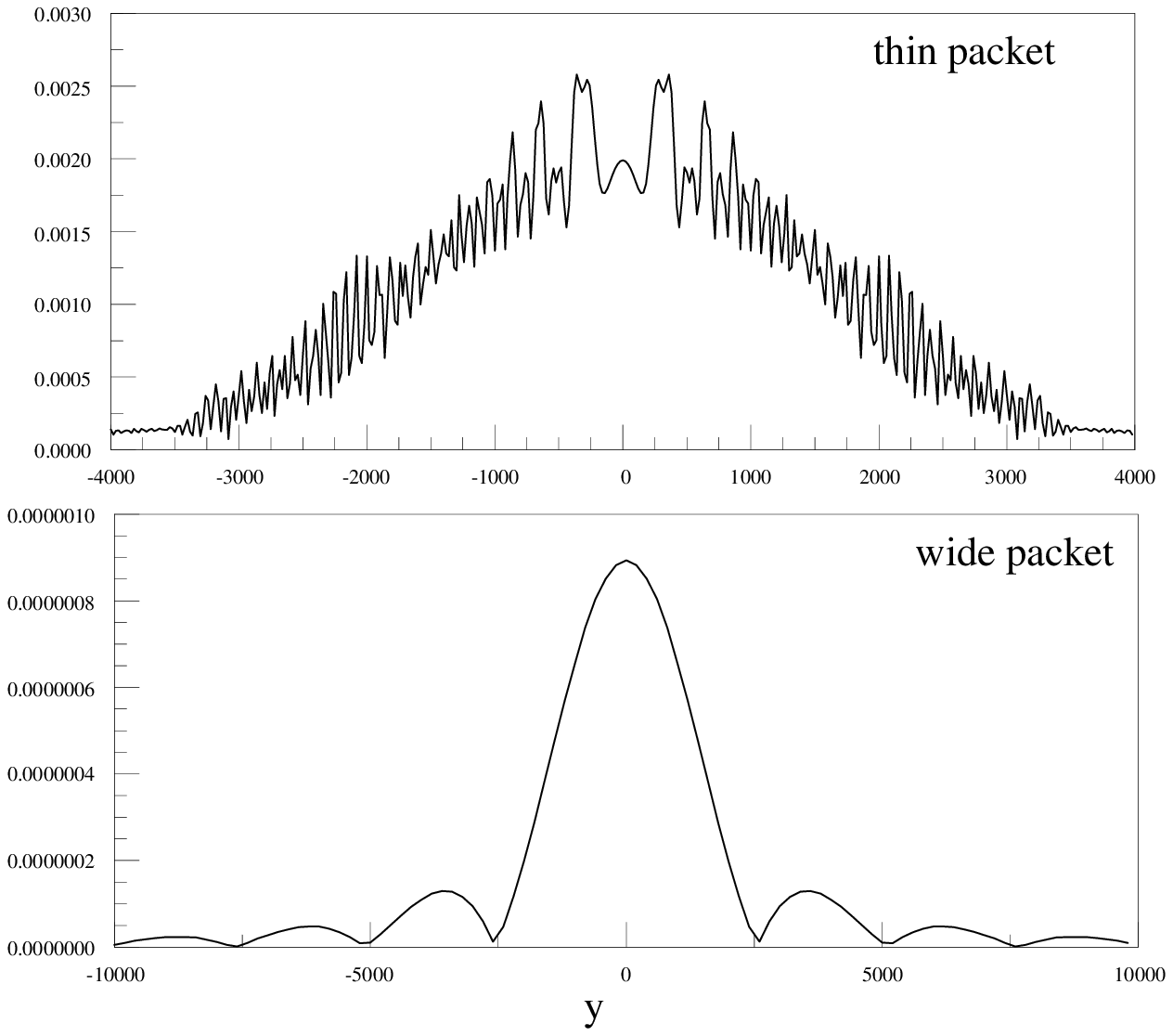}
\vsize=5 cm
\caption{\sl Calculated long time behavior of the diffraction
pattern of a thin as compared to a wide packet}
\label{fig15}
\end{figure}

The wide packet pattern resembles the diffraction pattern of a
plane wave, as expected. The thin packet produces a jagged structure
with many peaks, some of them quite pronounced over the background.

\section{\sl Summary}
In \ci{k1,k2,k3,k4}, the recently found effect of diffraction in space and time
with wavepackets, was addressed numerically and analytically. 
The diffraction pattern is evident in the backward zone in
one dimension and at large angles in two and three
dimensions.
In the present work, it was found that this backward effect
 appears also when packets scatter off slits.

The study of wavepacket scattering from slits 
has also uncovered a new diffraction effect.
This effect emerges quite unexpectedly with thin packets in
the forward direction.
A thin packet may be viewed as a mixture of many frequencies, the thinner
the packet, the wider the spectrum. Hence, a single central
peak is expected, and not any sidebands. The numerical
results contradict this prejudice.

The effect of \ci{k1,k2,k3,k4} was interpreted
as a product of the interference between incoming and reflected waves.
The forward diffraction pattern of the slits is here understood as
 resulting from the interference between incoming (transient) modes 
and the time dependent excitation of standing waves inside the slit
, thereby the name cavity mode diffraction.

The backward diffraction pattern appears as a hilly landscape along
the propagation axis, while the forward pattern appears across
this direction.

The results of the present work suggest that there is a 
need for a revision, and, a renewed 
effort to solve this paradigm of matter waves scattering.
On the experimental side, besides experiments with
 liquid Helium, as simulated in \ci{k3}, cold bose gas packets 
may provide an appropriate setting for the investigation of the
 new diffraction phenomena on slits.

\end{document}